\documentclass[showpacs,preprintnumbers,10pt,twocolumn]{revtex4}%
\usepackage{amssymb}
\usepackage{amsfonts}
\usepackage{amsmath}
\usepackage{graphicx}
\usepackage{dcolumn}
\usepackage{bm}
\usepackage{revsymb}%
\begin{document}
\title{Comment on \textquotedblleft Phase Control of Directed Diffusion in a
Symmetric Optical Lattice\textquotedblright}
\author{Ricardo Chac\'{o}n}
\affiliation{Departamento de F\'{\i}sica Aplicada, E.I.I., Universidad de Extremadura,
Apartado Postal 382, E-06006 Badajoz, Spain, and Instituto de Computaci\'{o}n
Cient\'{\i}fica Avanzada (ICCAEx), Universidad de Extremadura, E-06006
Badajoz, Spain}
\date{\today}
\maketitle

Motivated by investigation into the mechanisms that yield directed diffusion
in a symmetric periodic potential, Schiavoni \textit{et al}. [1] have studied
experimentally and numerically cold atoms in a one-dimensional dissipative
optical lattice where directed motion appears as a result of the breaking of
the system's temporal symmetry after applying a biharmonic phase modulation to
one of the lattice beams. In the accelerated reference frame, the atoms
experience a stationary optical potential together with an inertial force%
\begin{equation}
F\left(  t\right)  =\gamma\left[  A\cos\left(  \omega t\right)  +B\cos\left(
2\omega t-\phi\right)  \right]  , \tag{1}%
\end{equation}
where $\gamma$ is an amplitude factor, $A=1-B$, and the parameters
$B\in\left[  0,1\right]  $ and $\phi\in\left[  0,2\pi\right]  $ account for
the relative amplitude and initial phase difference of the two harmonics,
respectively. Commenting on their experimental results, the authors claim
that: \textquotedblleft By increasing $B$ from the zero value the atoms are
set into directed motion, and a maximum for the c.m.~velocity is reached for
$B\simeq0.5$, i.e., for about equal amplitudes of the even and odd
harmonics.\textquotedblright\ This statement has had the unfortunate
consequence that most of the subsequent published papers citing Ref.~[1] have
considered $B=1/2$ as the condition that maximizes the ratchet transport in
systems subjected to a biharmonic temporal force.

This Comment will question the above statement. I shall argue that the maximum
c.m.~velocity is reached for $B=1/3$ as predicted by the theory of ratchet
universality [2,3]. Indeed, it has been demonstrated for temporal and spatial
biharmonic forces that optimal enhancement of directed ratchet transport is
achieved when maximally effective (i.e., critical) symmetry breaking occurs,
which implies the existence of a particular universal waveform [2,3].
Specifically, the optimal value of the relative amplitude $B$ comes from the
condition that the amplitude of the odd harmonic must be twice that of the
even harmonic in Eq.~(1), i.e., $1-B_{opt}=2B_{opt}\Longrightarrow
B_{opt}=1/3$. This condition has been experimentally confirmed in the context
of a Bose-Einstein condensate exposed to a sawtooth-like optical lattice
potential [4]. Notice that this means that the contributions of the amplitudes
of the two harmonics to the directed motion of the atoms are \textit{not}
independent, which is solely taken into account in the estimate of the optical
pumping rate (escape rate) $\Gamma^{\prime}\propto\sin^{2}k\Delta z$, with
$\Delta z\propto A^{2}B$ being the displacement of the centre of oscillation
of the atoms in a potential well from the well centre [1], \textit{after} the
substitution $A=1-B$. In such a case, one obtains that $\Gamma^{\prime}%
=\Gamma^{\prime}\left(  B\right)  $ presents a single maximum at $B=1/3$ for
which the asymmetry between the escape rates towards the left and right wells
is maximal, and hence a maximal nonzero current of atoms is expected again for
$B=1/3$, as is indeed confirmed by the experimental results [1] (see Fig.~ 1).
Ratchet universality predicts that the strength of the nonzero current has the
functional dependence $\sim S\left(  B\right)  p\left(  \phi\right)  $ [3],
where $S(B)$ accounts for the degree of breakage of the shift symmetry
$F\left(  t+T/2\right)  =-F\left(  t\right)  $, while the $2\pi$-periodic
function $p\left(  \phi\right)  $ accounts for the degree of breakage of the
time-reversal symmetry $F(-t)=F\left(  t\right)  $ and presents two extrema at
the optimal values $\phi_{opt}=\left\{  \pi/2,3\pi/2\right\}  $. This
dependence on $\phi$ is indeed confirmed by the experimental results shown in
Fig.~2 of Ref.~[1]. Also, $S\left(  B\right)  $ presents features similar to
those of the impulse, $I\left[  F\right]  \equiv\left\vert \int_{T/2}F\left(
t\right)  dt\right\vert $, transmitted by the normalized version of the
biharmonic force (1), $F^{\ast}\left(  t\right)  $, for \textit{any} value of
$\phi$ (see Ref.~[3] for additional details). Figure 1 shows plots of the
function $S\left(  B\right)  $ for two limiting cases of the initial phase
difference: one of the optimal values $\left(  \phi=\pi/2\right)  $ and one of
the least favourable values $\left(  \phi=0\right)  $ [5]. These curves fit
the experimental data reasonably well, and present a single maximum at
$B=1/3$, as expected [2,3]. The importance of the results of Ref.~[1] is that
they provide a first experimental proof of ratchet universality in the context
of cold atoms in optical lattices.

The author thanks F. Renzoni and P. J. Mart\'{\i}nez for their help with the
experimental data recovery and interchanges about this issue.

\subsection{Figure Captions}

Fig.~1 (colour online). Velocity of the centre of mass of the atomic cloud for
$\phi=\pi/2$ (experimental data from Fig.~3 in Ref.~[1]; dots), and the curves
$S_{\phi=0}=1.95I\left[  F_{\phi=0}^{\ast}\right]  $ (dashed line) and
$S_{\phi=\pi/2}=6.6\left(  I\left[  F_{\phi=\pi/2}^{\ast}\right]  -1\right)  $
(solid line) as functions of the relative amplitude $B$ [see the text; Eq.~(1)].


\begin{thebibliography}{9}
\bibitem {5}The corresponding analytical expressions for $S_{\phi=0,\pi/2}$
can be obtained straightforwardly from the respective expressions in Ref.~[3]
after the substitution $\eta\rightarrow1-B$.
\end{thebibliography}
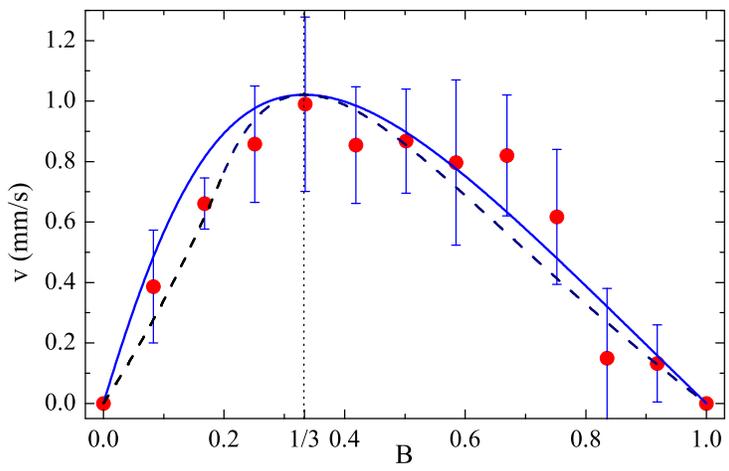
\end{document}